\documentclass[fleqn,usenatbib]{mnras}

\usepackage{newtxtext,newtxmath}

\usepackage[T1]{fontenc}
\usepackage{ae,aecompl}

\usepackage{graphicx}	
\usepackage{amsmath}	

\title[Tidally Excited Eigenmodes and Period Spacing]{A New Window to Tidal Asteroseismology: Non-linearly Excited Stellar Eigenmodes and the Period Spacing Pattern in KOI-54}

\author[Guo et al.]{Zhao Guo$^{1}$, Gordon I. Ogilvie$^{1}$, Gang Li$^{2}$, Richard H. D. Townsend$^{3}$, Meng Sun$^{3}$
\\
${1}$ Department of Applied Mathematics and Theoretical Physics, University of Cambridge, Cambridge CB3 0WA, UK \\
${2}$  Universit$\acute{e}$ de Toulouse, CNRS, CNES, UPS, Toulouse, France,\\
${3}$ Department of Astronomy, University of Wisconsin-Madison, Madison, WI 53706, USA\\
}

\date{Accepted XXX. Received YYY; in original form ZZZ}

\pubyear{2015}

\begin{document}
\label{firstpage}
\pagerange{\pageref{firstpage}--\pageref{lastpage}}
\maketitle

\begin{abstract}
We revisit the Tidally Excited Oscillations (TEOs) in the A-type, main-sequence, eccentric binary KOI-54, the prototype of heartbeat stars. Although the linear tidal response of the star is a series of orbital-harmonic frequencies which are not stellar eigenfrequencies, we show that the non-linearly excited non-orbital-harmonic TEOs are eigenmodes. By carefully choosing the modes which satisfy the mode-coupling selection rules, a period spacing ($\Delta P$) pattern of quadrupole gravity modes ($\Delta P \approx 2520-2535$ sec) can be discerned in the Fourier spectrum, with a detection significance level of $99.9\%$.  The inferred period spacing value agrees remarkably well with the theoretical $l=2,m=0$ g modes from a stellar model with the measured mass, radius and effective temperature. We also find that the two largest-amplitude TEOs at $N=90, 91$ harmonics are very close to resonance with $l=2,m=0$ eigenmodes, and likely come from different stars.

Previous works on tidal oscillations primarily focus on the modeling of TEO amplitudes and phases, the high sensitivity of TEO amplitude to the frequency detuning (tidal forcing frequency minus the closest stellar eigenfrequency) requires extremely dense grids of stellar models and prevents us from constraining the stellar physical parameters easily. This work, however, opens the window of real tidal asteroseismology by using the eigenfrequencies of the star inferred from the non-linear TEOs and possibly very-close-to-resonance linear TEOs. 
Our seismic modeling of these identified eigen g-modes shows that the best-matching stellar models have ($M \approx 2.20, 2.35 M_{\odot}$) and super-solar metallicity, in good agreement with previous measurements.
\end{abstract}




\section{Introduction}

Mode identification is a key step in asteroseismology and it relies heavily on pattern recognition in the Fourier spectrum of the oscillating star. In the asymptotic regime, gravity (g) modes are equally spaced in pulsation period \citep{Tas80}, and this period spacing pattern has been exploited in various kinds of 
self-excited g-mode pulsators such as $\gamma$ Dor stars \citep{Van16,Li20}, Slowly Pulsating B-stars (SPB)\citep{Pap17}, white dwarfs \citep{Alt10} and sub-dwarf B-stars \citep{Bar16}. The g-mode period spacing vs. period diagram has been used to infer the asymptotic period spacing values,  the near-convective-core rotation rates, the near-core-boundary mixing process of intermediate/massive stars, and the coupling between inertial modes and gravity modes \citep{Mor15, Oua17, Sai21}.

Gravity modes can also be excited externally by the tidal forcing from a binary companion. Significant advances have been made in tidal seismology after the discovery of the prototype heartbeat binary KOI-54 (HD 187091, KIC 8112039) by \citet[hereafter W11]{Wel11}.  Tidally Excited Oscillations (TEOs) have been observed in tens of heartbeat binary systems by the {\it Kepler} telescope \citep{Tho12,Kir16,Che20, Guo21}, and the {\it TESS} mission is unfolding more massive heartbeat binaries \citep{Jay19, Kol21}. Recently, nearly one thousand heartbeat stars in the OGLE survey have been reported by \citet{Wro21}.
Among the  heartbeat binaries, only a few systems have been studied in detail \citep{Guo17,Guo19,Guo20,Ham18,Ful17,Pab17,Jay21}.
It is realised that the frequency information is not so useful since they are essentially all orbital harmonics, i.e., a series of forcing frequencies. The harmonic TEO amplitude is very sensitive to the frequency detuning, i.e., the difference between the forcing frequency and an eigenmode frequency, which is difficult to obtain very precisely. This makes the seismic modeling very difficult, and extremely dense grids of stellar models are needed. \citet{Ful17} proposed a statistical approach to model the TEO amplitudes which takes into account this uncertainty in the detuning parameter.

In this work, however, we show that eigenfrequency information can be taken advantage of in the non-linearly excited g modes. We achieve this by studying the anharmonic TEOs of KOI-54 which are non-linear excited `daughter modes' undergoing three/multi-mode coupling. In Section 2, we briefly recap previous works on KOI-54. In Section 3, we show that there is a nearly equally spaced pattern in the pulsation periods of anharmonic TEOs. This period spacing ($\approx 2520-2535$ seconds) is in remarkable agreement with $l=2, m=0$ gravity modes of a stellar model whose mass, radius and effective temperature are consistent with the observationally-inferred values. In Section 4, we perform a grid-based modeling of these nonlinearly excited g modes and the two dominant linearly excited harmonics and explore the constraints we can put on the stellar parameters (mass, radius, and metallicity, etc). In the final section, we discuss the caveats of this work and suggest further works that can help to solve the remaining mysteries of this interesting binary system.

\section{Recap of previous works on KOI-54}

The discovery paper W11 measured the fundamental stellar parameters of KOI-54. The two main-sequence stars in the system have the same mass within $1\sigma$: $M_1=2.33\pm0.10, M_2=2.39\pm 0.12M_{\odot}$, and very similar effective temperatures: $T_{\rm eff1}=8500\pm 200$K, $T_{\rm eff2}=8800\pm 200$K. Only the radii are slightly different $R_1=2.20\pm 0.03, R_2=2.33\pm 0.03R_{\odot}$. Spectroscopy indicates a super-solar metallicity ([Fe/H]=0.4). The orbital parameters (period, eccentricity, and inclination) have also been determined: $P_{\rm orb}=41.805$d, $e=0.83$, $i=5.5^{\circ}$. Thus the orbital frequency is $f_{\rm orb}=0.02392$ day$^{-1}$.

Of particular interest are the orbital harmonic TEOs in this system, with the two dominant ones that are exact 90th and 91st orbital harmonics. This kind of stellar response is exactly within expectation: a forced harmonic oscillator oscillates at the forcing frequency and not at its intrinsic eigenfrequency. For eccentric binaries like KOI-54, the forcing frequencies are indeed a series of orbital-harmonic frequencies ($=Nf_{\rm orb}$, with $N$ being a positive integer).

Subsequent works took advantage of this perfect laboratory to study the effect of dynamical tides. \citet{Ful12} modeled the pulsation amplitude of the harmonic TEOs with the normal-mode decomposition approach. They did a detailed study of the nature of the 90th and 91st harmonics and suggested they can be naturally explained by resonance locking (RL), a phenomenon in which one pulsation mode is locked into resonance with the orbital frequency for a long time (when the evolution of the mode frequency and the forcing frequency have almost the same rate).

\citet{Bur12} (hereafter B12) also modeled the TEO amplitudes with the mode-decomposition formalism. In addition, since the non-adiabatic effect near the stellar surface is important in modeling the amplitudes, they also solved the linear, non-adiabatic forced oscillation equation, with rotation implemented in the traditional approximation. They also find that chance resonance of $l=2, m=0$ modes can explain the observed large amplitude of the 90 and 91 harmonics without invoking resonance locking.

\citet{Ole14} (hereafter O14) re-analysed the {\it Kepler} light curves with longer time coverage. They extracted and tabulated 70 harmonic TEOs and 50 anharmonic TEOs with amplitude larger than 0.7 $\mu$mag (their Table 2 and 3, respectively).
They also examined the TEO phases in great detail. The two dominant harmonic TEOs at $90f_{\rm orb}$ and $91f_{\rm orb}$ are found to be $l=2, m=0$ pulsations, so do the fifth and seventh largest harmonic TEOs at 72 and 53 harmonics. O14 discussed the anharmonics  arising from the non-linear three/multi-mode coupling in great detail. 

By using the data extracted in O14, we re-examine these nonlinearly excited modes and extend their analysis.

\section{Non-linearly Excited Anharmonic pulsations and gravity-mode period spacing}

We discuss the tidally excited oscillation in detail, and it is convenient to denote the frequencies by `$fN$', where $N$ is the frequency in units of orbital frequency ($N=f/f_{\rm orb}$). Thus the two dominant harmonic, linear TEOs are $f90$ and $f91$, and the largest-amplitude non-linear TEO (anharmonic) is $f22.419$. This can be seen in the Fourier amplitude spectrum of KOI-54 shown in Figure 1.
Note that in producing the data of Figure 1, O14 had already subtracted the equilibrium-tide contribution from the light curves. 

We distinguish between the orbital-harmonic TEOs (gray peaks) and the anharmonic TEOs (red peaks) in Figure 1. The harmonic TEOs are topped by green or blue filled circles and the corresponding orbital-harmonic number $N=f/f_{\rm orb}$. Those that are likely $l=2, m=0$ oscillations are marked by blue circles since they all have pulsation phases close to 0.25 or 0.75 ($\pm 0.02$ in units of  $2\pi$, see Table 2 in O14).  

To the linear order, the time-dependence of the stellar response is the same as the tidal forcing frequency which, in the case of eccentric orbits, are a series of orbital harmonic frequencies $Nf_{\rm orb}$.
In the framework of normal-mode decomposition, the stellar response can be expressed as linear combinations of stellar eigenfunctions, oscillating not at the eigenfrequencies but at the forcing frequencies $Nf_{\rm orb}$. 
The eigenfrequency comes into play in the resonance term, similar to a harmonic oscillator's response to an external forcing, and the oscillations have a Lorentzian-shaped response centred at the eigenfrequency. For a particular forcing frequency  $Nf_{\rm orb}$, the closest eigenfrequency $f_A$ (with the smallest detuning) dominates the amplitude of the dynamical stellar response, and the equilibrium-tide part of the stellar response comes from essentially all the eigenmodes, with the major contribution from the f-mode \citep[see][Fig. 1]{Wei12}.

Since the equilibrium-tide contribution has been removed, we can essentially think that only the single mode which is closest to the forcing frequency is oscillating, with the frequency of the forcing frequency (not its eigenfrequency). Because the linear stellar response are orbital harmonics and not eigenfrequencies, we cannot use frequency information in the seismic modeling. However, the anharmonic pulsations are likely generated by non-linear resonant mode coupling. They are actually eigenmodes of the star and contain useful information in their frequencies/periods.  Higher radial order g modes should satisfy the asymptotic relation and are nearly equally spaced in pulsation period. Here we show that this is indeed possible for tidally, nonlinearly excited g modes in KOI-54.

First, there are a few factors that help us to identify the pulsation modes. KOI-54 has a face-on orbit, with an orbital inclination of only $5.5^{\circ}$. We can expect the axisymmetric $m=0$ modes are more visible and $m \ne 0$ modes are strongly disfavoured observationally.
Harmonic TEOs, most likely linear, are most likely to be $l=2$. The anharmonic daughter modes though can be $l=1, 2, 3$, but the visibility of $l>3$ modes is significantly lower (actually while an $l=3$ mode is already very difficult to be seen in {\it Kepler} observations).

In Figure 1, a noticeable feature is a series of nearly equally spaced peaks in pulsation period, most notably between the pulsation period 0.4 and 1.0 day. The spacing is about 2500 seconds (5000s if missing one in the middle). These peaks are marked by the open brown circles and their spacings are labeled. This regular period spacing is more salient in Figure 2, where we have used two methods to search for regular spacing in the anharmonic TEOs. The upper panel shows the histogram of pairwise period differences \citep{Mac14}, and a significant over-density peak can be seen at $\approx 2520$ seconds. In the lower panel, we use the Kolmogorov-Smirnov (K-S) test to determine the significance of the regular period spacing. This method is widely used in the asteroseismology of compact stars \citep{Win91, Bar19}. The idea is that for a series of pulsation periods $\Pi_i$ (possibly regularly spaced with $\Delta \Pi$), we can calculate the deviation from the equal-spacing regularity $r_i=n_i - int(n_i)$, where $n_i=(\Pi_i - \Pi_{0})/\Delta \Pi$, $int(x)$ is the greatest integer less than $x$ and $\Pi_{0}$ is the shortest pulsation period. If the period series is random-distributed, $r_i$ should satisfy the uniform distribution from 0 to 1. Then we can use the K-S test quantify the difference between the actual $r_i$ of the data and the uniform distribution. As shown in Figure 2, a given period spacing is shown as a local minimum of $Q$, with ‘confidence level’ of $(1-Q)\times 100\%$. We find that $\Delta P \approx 2535$ sec at the $99.9\%$ significance level. Thus, both the upper and lower panels reveal a regular spacing of $\approx 2520-2535$ seconds. As will be detailed below, this spacing agrees remarkably well with the expected $l=2,m=0$ g-mode period spacing (See later Sections and Figure 4).

Note that in Figure 1, $f22.419$ is also in this $m=0$, brown-circled, regularly spaced peak series. It is also the largest-amplitude anharmonic pulsation, so likely having $l=1$ or $l=2$. In Figure 1, we have listed the $N$ values of anharmonic frequencies below the Fourier spectrum ($N=f/f_{\rm orb}$). It can be seen that four modes \textbf{are} coupled to $f22.419$:  $f68.582$, $f49.589$, $f30.587$ and $f26.579$. Thus they are more likely to have the same spherical degree $l$. These five modes are marked by the red squares.  These four daughter-mode pairs can pair up to form four parent-mode harmonics in three mode coupling ($f_A=f_a+f_b$): $f91\approx f22.419+f68.682$, $f72\approx f22.419+f49.589$, $f53\approx f22.419+f30.587$, and $f49\approx f22.419+f26.579$. The four parent modes they form are among the largest amplitude harmonic TEOs and they are all $l=2, m=0$ pulsations (topped by blue circles) as identified in O14. These five modes are listed in our Table 1.
 
Given an $l_A=2, m_A=0$ parent mode (e.g., $f_A=f91$ or $f72, f53$), the three-mode coupling selection rules can help to limit possible $(l, m)$ values of the daughters. If we reasonably assume $m_a=m_b=0$, only $(l_a, l_b)$ =(2, 2), (1, 1) or (1, 3) can satisfy the selection rules: $|l_a-l_b| \le l_A \le |l_a+l_b|$ and $(l_A+l_a+l_b) \ mod \ 2=0$.  B12 considered possible daughter modes of $f91$ within the range of $l_a,l_b \in 1-6$. They found that the optimal daughters are those with $l \in 1-3$. Among their listed optimal daughter pairs, the one with the smallest three-mode coupling threshold is the $(l_a, l_b) =$(2, 2) pair. Thus this combination is more likely to be observed. This supports the interpretation that the four modes coupled to $f22.419$ have $l=2, m=0$.

To better show the identified regular spacing of 2520s, an echelle diagram is constructed by using the anharmonic TEOs (the right panel of Figure 3). The symbols are scaled by the corresponding pulsation amplitudes. Modes with the same spacing form a vertical ridge in the echelle diagram, and indeed, the brown circled modes in Figure 1 all locate essentially in a vertical ridge. They all have a similar $P$ mod $\Delta P$ of about 2200 seconds. In total, we identified 16 candidates of $l=2,m=0$ eigenfrequencies. Detailed information on these equally-spaced modes is listed in Table 2. Note that the aforementioned four modes that coupled to $f22.419$ all locate essentially in a vertical ridge. We set the limit of the ridge as two vertical dashed lines. Our interpretation is that these 16 regularly spaced pulsations are a series of $l=2, m=0$ g modes. 

There are two frequency pairs with very small differencing: $f=57.675$, $f57.577$ and  $f49.731$, $f49.589$. Such close frequencies mean they cannot be both eigenfrequencies from one star, since their spacings are one order of magnitude smaller than the typical g-mode frequency spacing \footnote{If we include $l=1$ modes and $m \neq 0$ modes due to rotational splitting, this pair could still be from the same star.}. Our identified 16 g modes only include the larger-amplitude ones in the two pairs: $f57.577$, $f49.589$. In Table 2, lower-amplitude ones in the two pairs are shown in parentheses.



The left panel of Figure 3 shows the echelle diagram of the harmonic TEOs. We do not expect to see a vertical ridge unless the harmonic TEOs are very close to the eigenfrequency, i.e., with detuning much smaller than the period spacing $\Delta P$.  Most harmonics are, indeed, not vertically aligned and not within the vertical ridge delimited by the dashed lines. However, the $f90$ and $f72$ are inside the vertical ridge, and the $f91$ is also close to the ridge. These three frequencies happen to be the largest-amplitude orbital-harmonic $l=2, m=0$ TEOs (blue dotted peaks), and thus it is in fact expected that they should have smaller frequency detuning. The fact that they are indeed close to this vertical ridge (our identified $l=2, m=0$ eigenfrequencies) is encouraging. We will discuss further the detuning of the $f90$, $f91$ in the next Section.

Given the similarity of the two stars, we are not sure whether one star is pulsating or both stars are. Thus the pulsation spectrum may be a mixture of the spectra of both stars. Thus finding a g-mode period spacing from the same $(l, m)$ originating from one or two stars is very difficult. Our identified equally spaced eigenfrequencies mean that even in the mixed spectrum of two stars, there are still enough eigenmodes that have regular spacings. And these regularly spaced modes are likely from the same star.

In our 16 brown-circled $l=2, m=0$ modes, 9 can pair up to form a harmonic TEO, but not all their parent modes are visible. The parent modes are indeed not necessarily visible in the spectrum. 

A key question is whether the observed anharmonic pulsations could be self-excited instead of tidally excited.
Stars can indeed show both tidally excited g modes and self-excited g modes \citep[][KIC4142768]{Guo19}. Luckily, KOI-54 locates in the gap between the SPB instability strip \citep{Pam99} and the $\gamma$ Dor instability strip \citep{Xio16}. Its masses of $2.33M_{\odot}$ and $2.35M_{\odot}$ are too massive to be a $\gamma$ Dor star and not massive enough to be a SPB star. In fact, our non-adiabatic calculations in Section 4 show no unstable g modes. Thus it is reasonable to assume that all the anharmonic pulsations are of tidal origin, as opposed to self-excitation.




Our interpretation of a series of $l=2,m=0$ modes spaced by about 2520 seconds can naturally explain the daughter modes since the series include almost all the largest amplitude daughters. 

We can have further constraints on the mode identification if we consider the spatial overlap of the eigenfunctions of daughter modes. The coupling coefficient between daughters is only significant if they have good spatial overlap. This requires that the difference between the daughter radial orders be less than the parent's, i.e., $|n_a-n_b| \lesssim n_c$ \citep[see, e.g.,][Fig.12]{Wei12}. If the modes in a triplet are short wavelength g-modes (for which eigenfrequencies satisfy $\omega_{nl} \propto l/n$), then the small detuning ($f_a + f_b \approx f_c$) and spatial overlap conditions together imply that the daughter frequencies satisfy $|f_a/f_b - f_b/f_a| \lesssim 1$. As shown in the last column of Table 1, this condition is only satisfied for the last two pairs listed. However, it requires further study to see whether this criteria applies strictly to the radial orders in our case ($n \sim 10-25$ v.s. $n\sim 500$ in the solar g-modes of Weinberg et al.\ 2012). Our preliminary calculation seems to show less drastic drop for $n_c > |n_a-n_b|$. Also, the threshold amplitude of the parent mode for the three-mode-coupling  $S_a$ \citep[eq.38]{Bur12} depends both on the coupling coefficient $\kappa_{abc}$ and the frequency detuning $\delta \omega$. Daughter mode pairs with smaller $\kappa_{abc}$, once compensated with smaller $\delta \omega$, can still reach the threshold for three-mode-coupling. Thus the first two daughter pairs in Table 1 could still be the daughter modes of f91.

 The $m=\pm 1$ modes also satisfy the selection rules for mode coupling. If these $m=+1,-1$ modes are present, pulsations with same $l$ and $m$ are also equally spaced in periods, with a period spacing of $\sim 2520$ sec for $l=2$ modes, and $\sim 4370$ sec for $l=1$ modes . In fact, the K-S test in Figure 2 shows a possible period spacing of 4760 seconds (although at much lower significance level), which is not too far away from the $l=1$ mode period spacing, which is $\approx \sqrt{3}$ times of the $l=2$ counterpart.  We cannot rule out the possibility that the anharmonic frequencies may contain some $l=1$ modes. Rotational splittings may also be present, with 
a frequency spacing of $\delta f =(1-C_{nl})f_{\rm rot}$, with $C_{nl} \approx 0.5,0.16$ for $l=1,2$ modes, respectively. These are regular frequency spacings not regular period spacings. Given the projected rotational velocity $v\sin i=7.5$ km/s and orbital inclination $i=5.5$ deg (W11), we obtain a rotational frequency $f_{\rm rot}=0.70$ d$^{-1}$.  Thus we expect to find rotational splittings of $\delta f=0.35$ d$^{-1}$ for $l=1$ modes; and $\delta f=0.59$ d$^{-1}$  for $l=2$ modes. However, a search for regular frequency spacing in the anharmonic TEOs yields no salient significant frequency spacings.

\section{Comparison with Stellar Models and Seimic Modeling}

After identifying a possible $l=2, m=0$ g-mode period spacing, we proceed to compare the observed eigenfrequencies with those from stellar models. 

We evolve a series of non-rotating stellar models with the MESA(v8118) stellar evolution code \citep{Pax11,Pax13,Pax15,Pax18}.  
We search in the following effective-temperature/radius/mass parameter space: $T_{\rm eff} \in (8100, 9200K)$,  $R \in (2.0-2.4R_{\odot})$  $M \in (2.00-2.60M_{\odot})$. For both stars, these parameter ranges cover the $\pm 2\sigma$ of the observed $T_{\rm eff}$, $\pm 3\sigma$ range of the observed radii,
 and $\pm 2\sigma$ range of the observed masses.
We used a slightly larger search-box for stellar radius. The reason is that KOI-54 in not an eclipsing system and the inference on radius (based on relative radius $R/a$ for an eclipsing system) is rather limited. The radius errors ($\sigma_ {R_{1,2}} =0.03R_{\odot}$) derived in W11 are probably underestimated.

The adopted step size in mass is $0.05M_{\odot}$ and our maximum time step in the evolution is set to $8 \times 10^6$ yrs. We also explore the effects of convective-core overshooting and metallicity. Our calculations adopt the exponentially decaying 
overshooting \citep{Her00}, and covers the parameters 
$f_{ov}= 0.00$ and $0.02$. We adopt the zero point for the overshooting function $f_0=0.001$. We fix the convective mixing length parameter $\alpha_{\rm MLT}$ to 1.8 and adopt the OPAL opacity tables \citep{Igl96} and default MESA equation-of-state.

The observed metallicity is super-solar: $[Fe/H]=0.4\pm 0.2$. Following Nsamba 2019 and using the `gs98' solar mixtures \citep{Gre98} with an initial helium abundance $Y_0=0.2484$, we find the [Fe/H]=0.4 corresponds to metal mass fraction $Z=0.035$. We thus calculate stellar models with $Z=0.02$, $0.03$, $0.04$.

For each MESA model within the observed radius error box $R\in (2.0-2.4) R_{\odot}$, we calculate the non-adiabatic eigenfrequencies\footnote{The effects of rotation on the eigenfrequencies are not taken into account, and this is a reasonable approximation since we are dealing with $m=0$ modes.} with the GYRE(v5.0) oscillation code \citep{Tow13,Tow18}. The pulsation frequencies of 16 observed $l=2, m=0$ g modes are compared with the calculated eigenfrequencies and the root mean square errors (RMSE) are calculated. It is defined as $RMSE=\sqrt{ \sum{(f-f_{\rm obs})^2}/N}$ (in units of day$^{-1}$ in the figure) and is a measure of, in an average sense, how well we can match the observed frequencies. Figure 4 shows the result for one of our best-fitting models with $M=2.20M_{\odot}, Z=0.03, f_{ov}=0.02$, and we call it our `baseline model'. The upper panel shows the evolution of g-mode eigen-periods (open circles) as a function of stellar radius. The identified $l=2,m=0$ g modes are indicated by the horizontal brown lines. We also include a likely $l=2, m=0$ mode with $f=63.067f_{\rm orb} $(yellow line); this daughter mode has a large amplitude, and is also close to the vertical ridge in the echelle diagram. It is likely that this mode belongs to the series. It deviates to the left of the vertical ridge probably due to mode trapping. Indeed, in the Period Spacing vs Period diagram, we can usually see these trapped g modes having smaller $\Delta P$ and deviate from the equal spacing. 

The two dominant orbital harmonic pulsations at $N=90$ and $91$ (blue lines) are also shown. The reason is that these two pulsations are believed to be very close to eigenmodes (resonances). In the TEO amplitude modeling of B12, they found that in order to achieve the observed large amplitudes of $f90$ and $f91$, a chance resonance with $l=2, m=0$ modes with a frequency detuning $df \approx 0.01f_{\rm orb}=0.0002$ day$^{-1}$ is required.


First, we show that the observed period spacing $\Delta P \approx 2520$s is in nice agreement with the expected $l=2,m=0$ g modes of a star with the measured mass, radius, and effective temperature in W11. This can be seen obviously in the upper panel of Figure 4. The theoretical pulsation periods from GYRE can be compared with the horizontal brown/yellow lines. In the lower panel, we show the frequency matching result, i.e., the RMSE as a function of stellar radius, $T_{\rm eff}$ and age. The five best-matching models are marked by the blue squares. The best-matching model (indicated by the filled blue square) has $R = 2.19R_{\odot}$ and $T_{\rm eff} \approx 8500$K and the corresponding stellar age is 400 Myr.

When the stellar mass varies, we find that $M=2.00$ and $2.05M_{\odot}$ models have strong avoided crossings, since for a lower-mass model to have the same radius, the star must be more evolved. And the g modes appear not as regular as those in Figure 4. 
Models with $M>2.4M_{\odot}$ cannot match the observed $R$ and $T_{\rm eff}$ constraints.
Thus the regular appearance of g modes matches the mass range of $M=2.1-2.3M_{\odot}$, which have only mild avoided crossings and do not deviate strongly from the regular period-spacing pattern.

By using models with $M$ from 2.00 to 2.60 $M_{\odot}$ with $\Delta M=0.05M_{\odot}$ (fixed $Z=0.03$, $f_{ov}=0.02$), we explore whether we can match the $f90$ and $f91$ in Figure 5.  We find that for models within the observed $R$ and $T_{\rm eff}$ box (dashed lines), $f90$ and $f91$ cannot be matched simultaneously. This is because the eigenfrequency spacing of the models is larger than the orbital frequency. The lower panel shows the models with absolute frequency difference $|f-f90|<0.004$ d$^{-1}$ (in green) and $|f-f91|<0.004$ d$^{-1}$ (in blue). The upper panel uses a smaller threshold of 0.0008d$^{-1}$. This small detuning value is in the same order of magnitude as that required to ensure $f90$ and $f91$ have the observed large amplitudes ($\approx 0.2-0.3$ mmag).

We also show the mean absolute frequency difference result for the 16 identified $l=2, m=0$ g modes. Models having low RMSE (or denoted as $|f-f(m=0)|$ shown in Figure 5) are indicated by the brown circles, with the lower panel adopting a larger threshold $0.02$d$^{-1}$ and the upper panel a smaller threshold $0.015$d$^{-1}$.

Note that f91 is the parent mode of several daughters in the $l=2,m=0$ g mode series, so we expect $f91$ and the 16 $l=2,m=0$ g modes are from the same star. In our modeling shown in Figure 5, there are indeed models having both the brown circles and blue filled circle marks, these models are best candidates for the inferred stellar parameters of KOI-54. The top panel, with a smaller threshold, shows that models with $M=2.20M_{\odot}$ and $R\approx 2.2-2.3R_{\odot}$ can match both f91 and the $l=2, m=0$ g-mode series reasonably well.

In both thresholds, $f90$ and $f91$ cannot be matched simultaneously, i.e., there are no models with both filled-blue and filled-green marks. $f90$ and $f91$ must come from two different stars. This fact was also recognised in B12. The top panel indicates that $f90$ can be matched to $<0.0008$d$^{-1}$ for the higher-mass model with $M=2.35M_{\odot}$.  A binary model with $M_1=2.20M_{\odot}$ and $M_2=2.35M_{\odot}$ can naturally explain the observed TEOs: the former can match the 16 $l=2, m=0$ eigenmodes and $f91$, and the latter has an eigenmode resonating with $f90$.  Given the similarity of the two stars, it would be surprising that only one star shows TEOs and the mode-coupling and the other does not.

Note that when fitting the $l=2,m=0$ g-mode frequencies, our grids of models cannot make the mean of the absolute theoretical-observed frequency difference $|f-f(m=0)|$ smaller than $0.01$d$^{-1}$. It is possible that the identified series of $l=2, m=0$ modes is contaminated with signals from the other star. In the mixed Fourier spectrum of both stars, the period spacing pattern is compromised and the comparison with one stellar model is thus not very successful. However, because of the similarity of the two stars, the observed g-modes still show a regular period spacing.


In Figure 6, we compare the eigenfrequencies of models of different metallicities and convective-core overshooting with the observed 16 $l=2, m=0$ g modes.
As shown in the right panel, we find solar metallicity models ($Z=0.02$) generally perform worse than super solar metallicity models ($Z=0.03$), having larger RMSE (except for a narrow radius range of $2.26-2.40R_{\odot}$ for $M=2.10M_{\odot}$ models). This is in agreement with the spectroscopic measurements in W11 who found $Z\approx 0.035$.

The left panel shows the result for the convective-core overshooting. Both $f_{ov}=0.00$ and $f_{ov}=0.02$ models can match the frequencies reasonably well, and they have essentially similar performance. 






\begin{figure*}
	\includegraphics[width=2.0\columnwidth,angle =0]{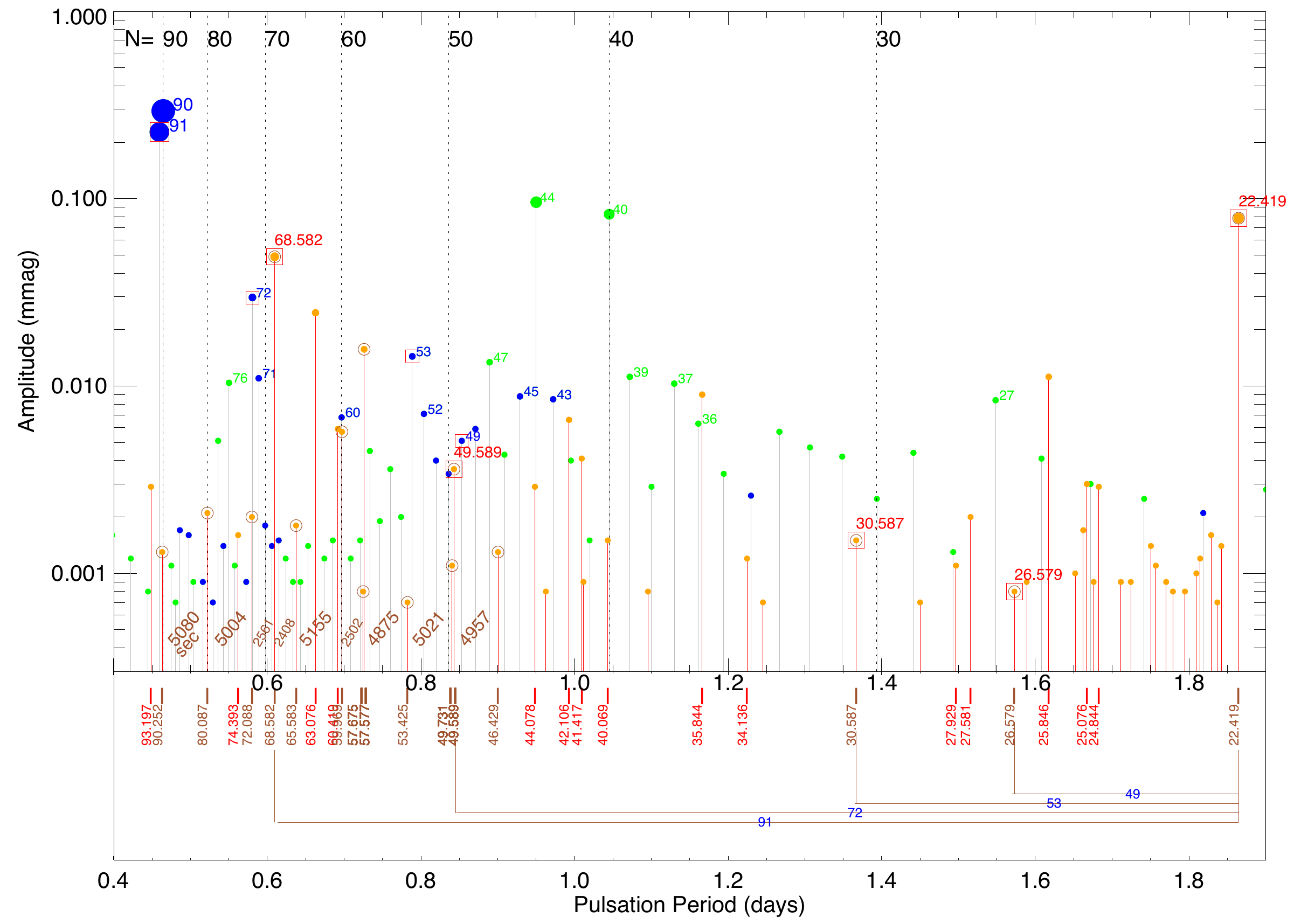}
    \caption{Distribution of pulsation amplitudes down to $1\mu$ mag. The symbol sizes are scaled according to the pulsation amplitudes. Orbital-harmonic pulsations are shown as gray peaks and those having $m=0$ phases are topped by blue circles. Anharmonic pulsations are indicated by the red peaks. A series of regularly spaced peaks are marked by the brown circles and they are candidates of $l=2,m=0$ eigenmodes. The largest anharmonic pulsation at 22.419 times the orbital frequency is labeled by a red square, and it couples with four other anharmonic frequencies indicated by the same symbol. The coupling is indicated by brown lines connecting these five frequencies. The orbital harmonic numbers $N$ of the corresponding parent modes are shown on these lines. Some orbital harmonic pulsations and most of the anharmonic frequencies are labeled by their $N=f/f_{\rm orb}$, either on the top or at the bottom.}
    \label{fig:1}
\end{figure*}

\begin{figure*}
	\includegraphics[width=1.9\columnwidth,angle =0]{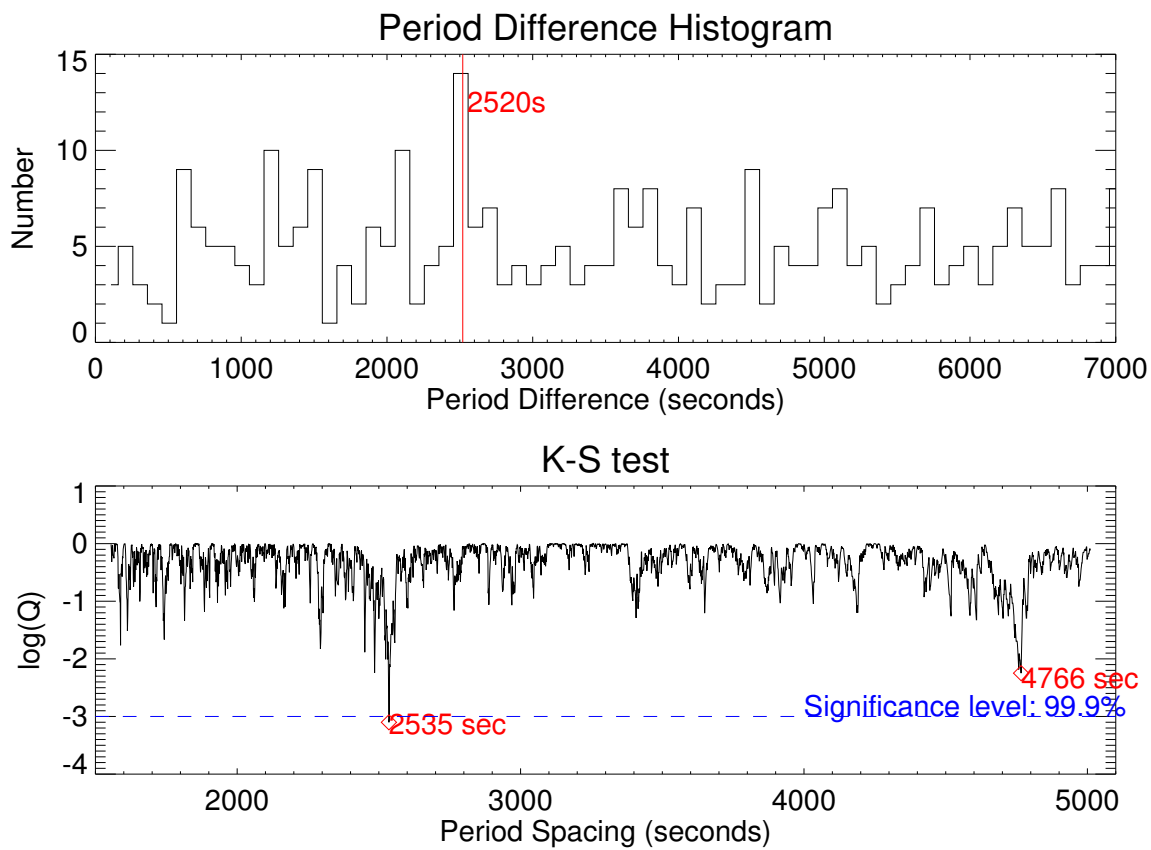}
    \caption{\textbf{Upper}: Period difference histogram of the anharmonic pulsations. \textbf{Lower}: Kolmogorov-Smirnov (K-S) test of the anharmonic pulsation periods. In both diagrams, a salient period spacing of $\approx 2520-2535$ seconds is revealed. The K-S test statistic $Q$ indicates that the regular period spacing at $\Delta P \approx 2535 $ sec is significant at the $99.9\%$ confidence level. }
    \label{fig:2}
\end{figure*}

\begin{figure*}
	\includegraphics[width=2.2\columnwidth,angle =0]{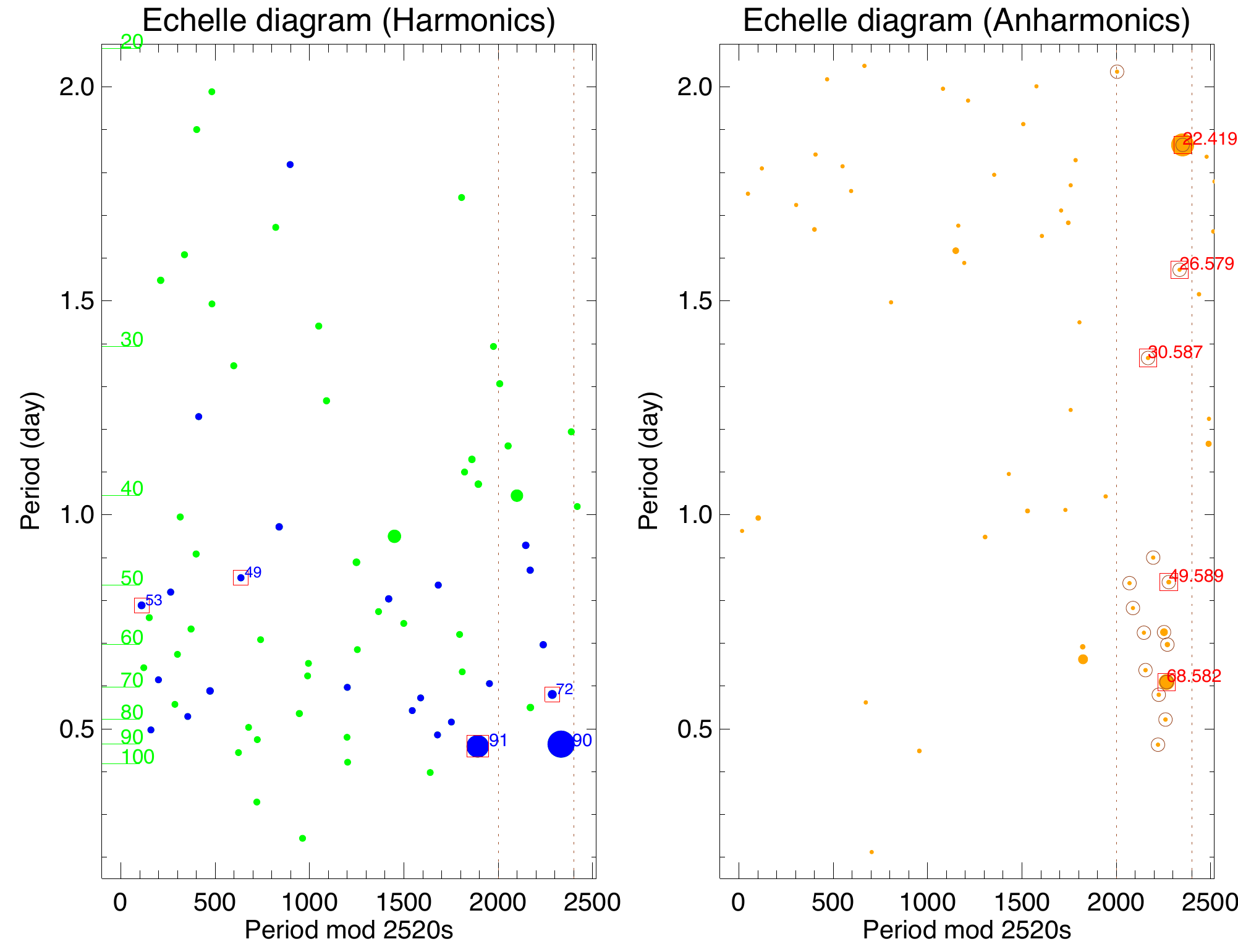}
    \caption{Echelle diagram of the orbital harmonic pulsations (left) and the anharmonic pulsations (right). The largest-amplitude anharmonic pulsation at $f/f_{\rm orb}=22.419$ is labeled by the red square, together with the four modes it couples. The parent modes of these four daughter-mode pairs are also marked by the red square in the left panel, with their $N=f/f_{\rm orb}$ labeled. }
    \label{fig:3}
\end{figure*}

\begin{figure*}
	\includegraphics[width=2.0\columnwidth,angle =0]{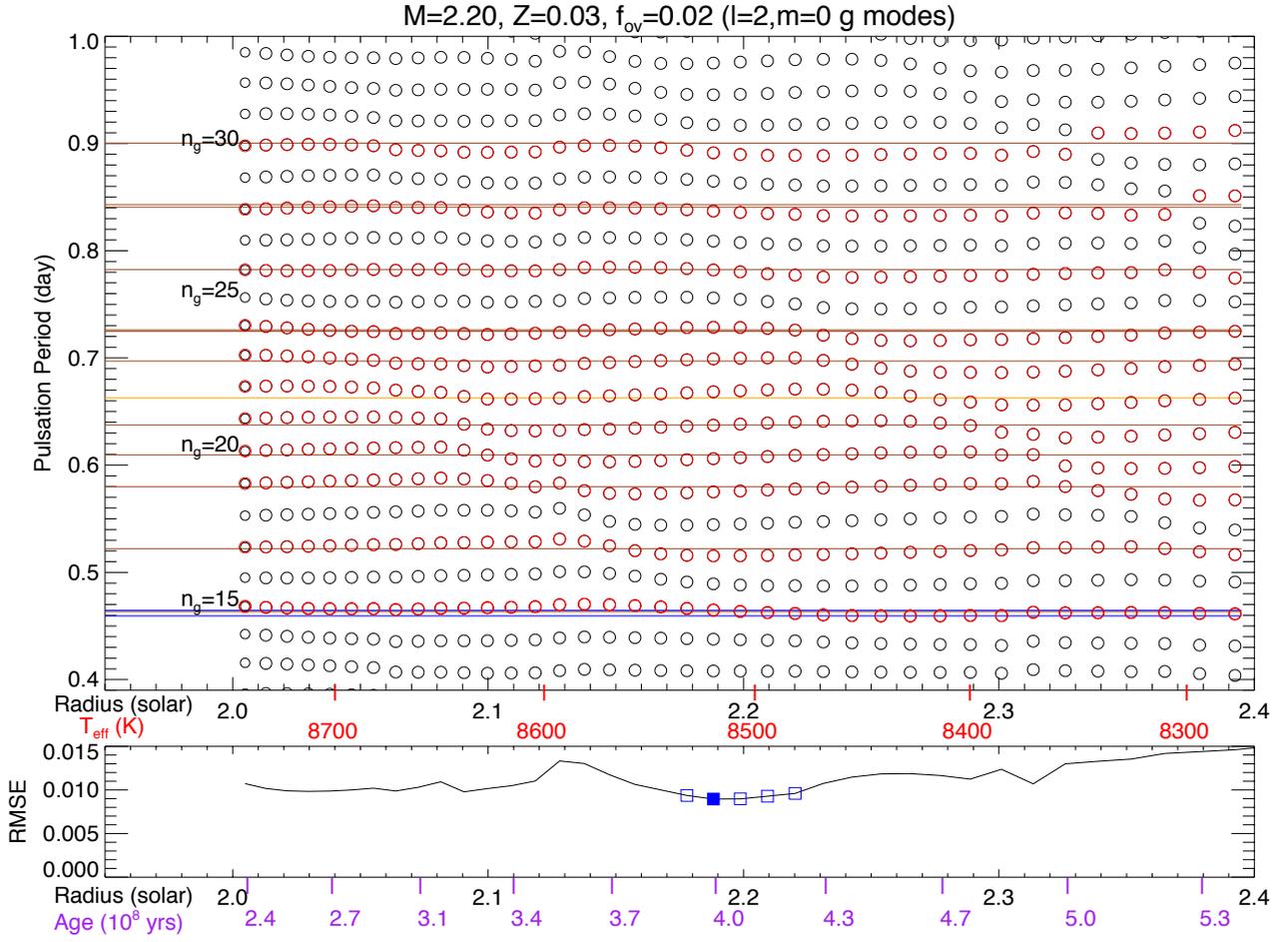}
    \caption{\textbf{Top}: The eigenmode periods for a stellar model of $M=2.20M_{\odot}, f_{ov}=0.02, Z=0.03$ is shown as a function of stellar radius, effective temperature and age. These eigenmode periods (circles) can be compared with the 16 observed $l=2, m=0$ g modes (shown as horizontal lines and they are marked by the brown circles in Figure 1, 2). The two dominant harmonic pulsations at 90 and 91$f_{\rm orb}$ are also shown (blue lines). \textbf{Bottom}: Frequency matching to the 16 observed g modes. The Root Mean Square Error (RMSE) is plotted for each model.}
    \label{fig:4}
\end{figure*}

\begin{figure*}
	\includegraphics[width=1.6\columnwidth,angle =90]{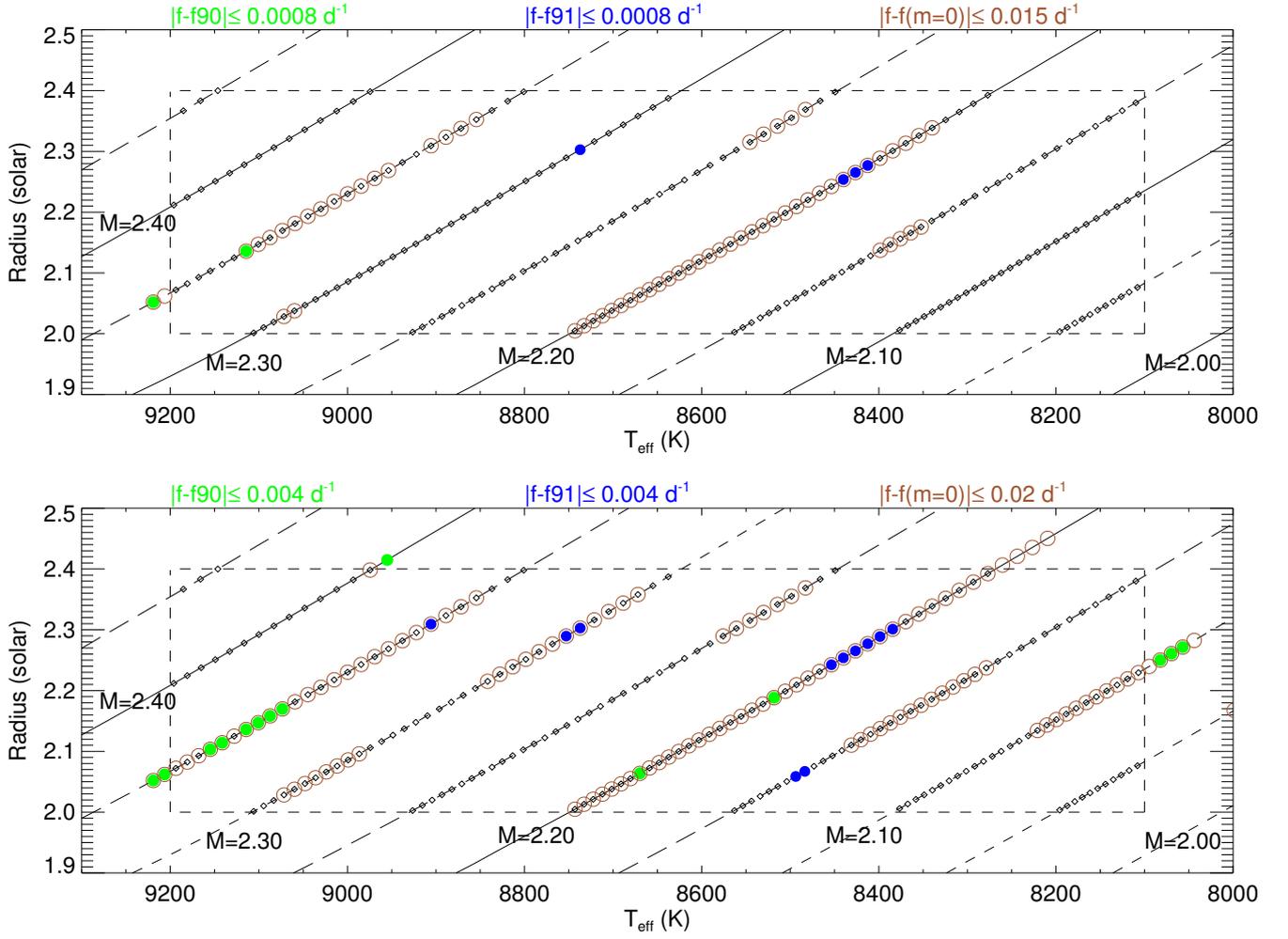}
    \caption{Evolutionary tracks on the $T_{\rm eff}$-Radius plane. The observed $T_{\rm eff}$-R box is indicated by the dashed boundaries. The eigenfrequencies inside the box are calculated, and those that satisfy certain conditions are labeled. Three cases are considered: the mean absolute frequency difference f(model)-f(observed) for the observed harmonics:  f91 (blue), f90 (green) and the 16 observed $m=0$ g modes (brown). The upper and lower panels use different thresholds, in the frequency units of day$^{-1}$. }
    \label{fig:5}
\end{figure*}

\begin{figure*}
	\includegraphics[width=1.6\columnwidth,angle =90]{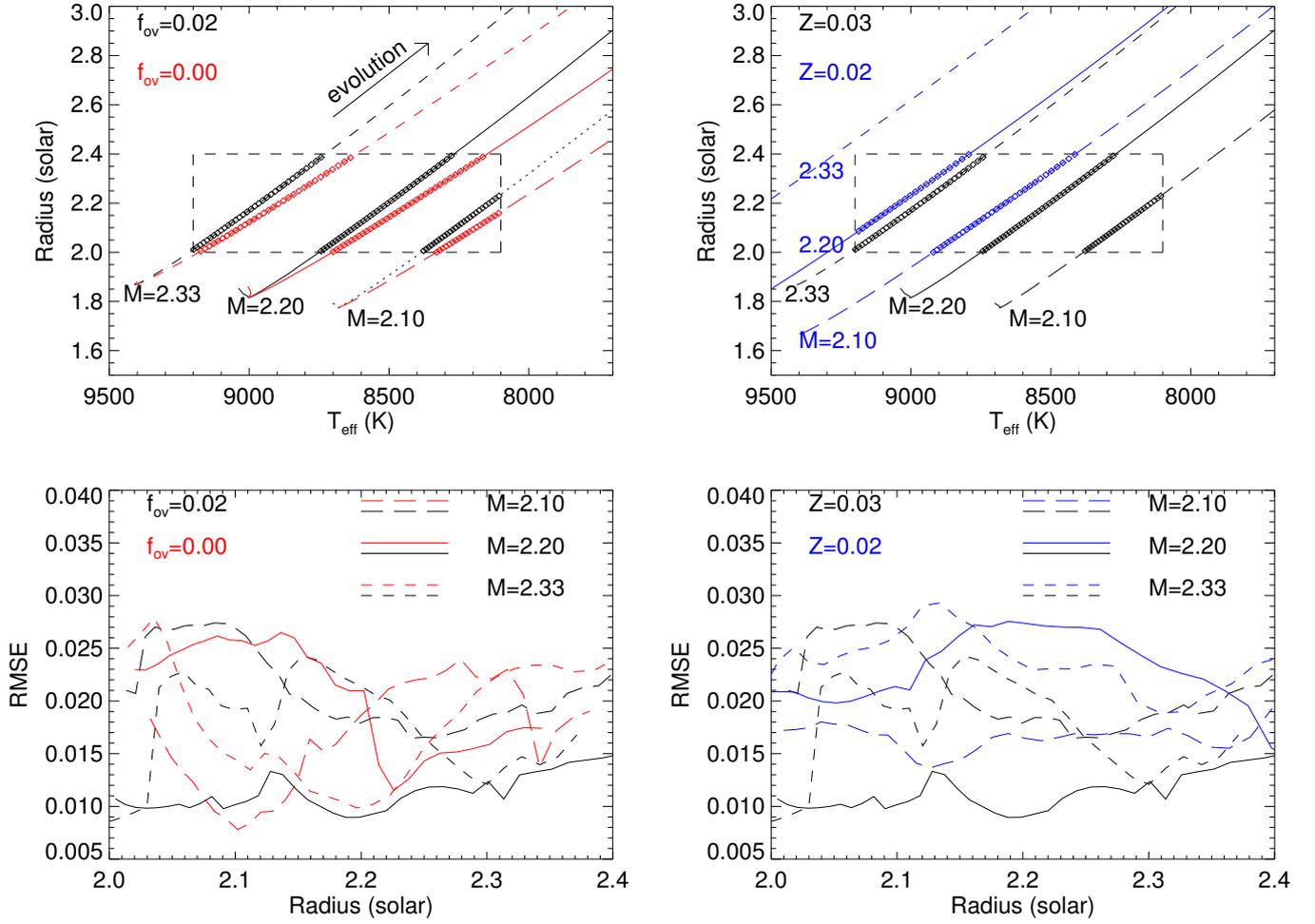}
    \caption{\textbf{Upper}: Evolutionary tracks on the Radius-Effective temperature plane. The effects of different convective overshooting (left panel) and metallicity (right panel) are shown. The observed $R-T_{\rm eff}$ box is indicated by the dashed lines and models inside are calculated for their eigenfrequencies. \textbf{Lower}:  Root Mean Square Errors from matching the 16 observed $l=2, m=0$ eigen g modes with eigenfrequencies from stellar models. }
    \label{fig:6}
\end{figure*}

\begin{table}
	\centering
	\caption{Four daughter modes coupled to f22.419.}
	\label{tab:example_table}
	\begin{tabular}{lcccr} 
		\hline
		$N(f_a)$ & $N(f_b)$ & $N(f_a+f_b)$ & N & $ |f_a/f_b- f_b/f_a|$\\
		\hline
		68.582 & 22.419 & 91.001 & 91 & 2.7322\\
		49.589 & 22.419 & 72.008 & 72 & 1.7598\\
		30.587 & 22.419 & 53.006 & 53 & 0.6314\\
		26.579 & 22.419 & 48.998 & 49 & 0.3420 \\
		\hline
	\end{tabular}
\end{table}


\begin{table}
	\centering
	\caption{Identified $l=2, m=0$ g modes and the two largest harmonic TEOs f90 and f91.}
	\label{tab:example_table}
	\begin{tabular}{lcccr} 
		\hline
		Frequency  & N & Amplitude & Period & P mod $\Delta$P\\
		(day$^{-1}$) &($f/f_{\rm orb}$)   & (mmag) & (day) & (sec)\\
		\hline
       2.158883 &        90.252 &     0.0013 &       0.4632025 &        2221\\
       1.915717 &        80.087 &     0.0021 &       0.52199777 &        2261\\
       1.72439 &        72.088 &     0.0020 &       0.57991522 &        2225\\
       1.640532 &        68.582 &      0.0490 &       0.60955836 &        2266\\
       1.568781 &        65.583 &     0.0018 &       0.63743760 &        2155\\
       1.434493 &        59.969 &     0.0057 &       0.69711041 &        2270\\
       (1.379613) &        57.675 &    0.0008 &       0.72484095 &        2146\\
       1.377281 &        57.577 &      0.0157 &       0.72606825 &        2252\\
       1.277955 &        53.425 &    0.0007 &       0.78250017 &        2088\\
       (1.189604) &        49.731 &     0.0011 &       0.84061587 &        2069\\
       1.186201 &        49.589 &     0.0036 &       0.84302745 &        2278\\
       1.110606 &        46.429 &     0.0013 &       0.90040933 &        2195\\
      0.731666 &        30.587 &     0.0015 &        1.3667438 &        2167\\
      0.635773 &        26.579 &    0.0008 &        1.5728884 &        2338\\
      0.536266 &        22.419 &      0.0787 &        1.8647462 &        2354\\
      0.491229 &        20.536 &     0.0010 &        2.0357104 &        2005\\
       1.508813 &        63.076? &      0.0246 &       0.66277266 &        1824\\
       \hline
       2.176809 &        91.002 &       0.2277 &       0.46466667 &        1891\\
\hline
\hline
       2.152855 &        90.000 &       0.2942 &       0.46466667 &        2347\\
		\hline
	\end{tabular}
\end{table}

\section{Discussion and Conclusions}

Based on previous works of B12 and O14, the main contribution of this work lies in finding the period spacing pattern in the nonlinearly excited eigenmodes.
This method of mode identification has been used extensively for self-excited g-modes but never for tidally excited modes. This promising method can be potentially applied to tens of {\it Kepler} heartbeat binaries with TEOs, as well as those from other space missions.
We also show that the linearly excited TEOs, i.e., orbital harmonic pulsations, can also be used as eigenmodes if the frequency detuning is sufficiently small.

However, there are caveats in this work, and lots of uncertainties still remain for KOI-54.

We did not model the amplitudes of orbital-harmonic TEOs. This was done in B12 and it is found that non-adiabatic calculations and dense stellar models are required. We also did not consider the effect of rotation in our eigenfrequency calculations since $m=0$ mode frequencies are unaffected by rotation to the first order. Rotation does play an essential role in shaping the appearance of the pulsation Fourier spectrum (B12).

Now that possible eigenfrequencies have been identified, 
it is desirable to perform a thorough modeling of the amplitudes and phases of harmonic TEOs including all the subtle effects mentioned above, together with our newly identified $l=2, m=0$ g modes. The large uncertainty in the frequency detuning can now be significantly reduced.

There are also lots of unidentified anharmonic TEOs and they could be gravity modes of $l=1-3$. By calculating the mode-coupling thresholds for various of daughter-mode pair combinations (different $l, m, n$), it is possible
to narrow down their mode identifications. Also, the $m \ne 0$ harmonic TEOs are mostly unidentified. O14 did a preliminary identification by using their phases, and they have possibly $l=1, 2$ or even $l=3$. 

There are possible $l=2,m=0$ modes that we have missed. They may not locate in the vertical delimited region in the echelle diagram. One piece of evidence is from $f91$. B12 and O14 showed that it has four daughter pairs: ($f22.419$, $f 68.582$), ($f60.419, f30.587$), ($f49.589, f41.417$), and ($f63.076, 27.929$). We already identified the first pair as $l=2,m=0$ modes, and $f49.589$ and possibly $f63.076$ are also in our identified series. It is thus reasonable to assume that some of the remaining daughter modes in these pairs also belong to the $l=2,m=0$ mode series.

It is expected that additional modeling effort could potentially further constrain the stellar parameters of both stars in KOI-54.

We are studying pulsations with amplitudes down to 1 $\mu$mag. Such low amplitudes may raise questions such as: Are we over-interpreting the data?  Does the data reduction process in O14 produce any fictitious frequencies? Although we cannot completely rule out these possibilities, it is hard to believe that all the convincing evidence we find in the data is purely due to coincidences. Besides, our work, apart from being incremental progress to the characterisation of KOI-54,   points out a new piece of information:  the direct usage of anharmonic TEOs in tidal asteroseismology.

Throughout the paper, we have assumed the anharmonic TEOs oscillate at exactly the linear eigenfrequency. But at some level nonlinear interactions between modes shift the daughter oscillation frequencies from their linear value \citep[see, e.g.,][Appendix C]{Kum96}. This subtle effect is an important mechanism for weakly nonlinear modes and could be very useful to link the theory with observations. It is also interesting to check the frequency detunings in the mode triplets and to compare them with the theoretical parametric instability width \citep{Kum96}. We plan to perform further studies in a separate paper.

We also did not study the amplitude/phase variation of the TEOs. O14 showed that the amplitudes of $f90$ and $f91$ are decreasing with time. It is possible to extract this information for all TEOs. And in principle, we can model the linear and nonlinear TEO amplitude variation in the  `amplitude equations (AE)' framework \citep{Wei12}. This would involve calculating the mode-coupling coefficients and integrating the AEs accordingly \citep{Guo20}. The nonlinear TEOs can settle into an equilibrium state, and their amplitude ratios are proportional to their corresponding damping rates. Of course, they can also have different behaviors such as limit cycles or even chaos. There remains lots of room for the study of nonlinear seismology.

\section*{Acknowledgements}

We thank Hiromoto Shibahashi for his enlightening questions in the PIMMS meeting. We are in debt to the anonymous referee for very insightful comments and suggestions, which greatly improved the quality of this paper. This research was supported by STFC through grant ST/T00049X/1. GL acknowledges support from the project BEAMING ANR-18-CE31-0001 of
the French National Research Agency (ANR) and from the Centre
National d’Etudes Spatiales (CNES). RHDT and M Sun acknowledge support from NSF grant ACI-1663696 and NASA grant 80NSSC20K0515.
\section*{Data Availability}
The data underlying this article were accessed from O14. The derived data generated in this research will be shared on reasonable request to the corresponding author.


\bibliographystyle{mnras}
\bibliography{koi54} 











\end{document}